\documentclass[
aps,
amsmath,amssymb,
jvst,
reprint,
]{revtex4-2}

\usepackage{xcolor}
\usepackage{caption}
\usepackage{subcaption}
\usepackage{graphicx}
\usepackage{ulem}
\usepackage{mathrsfs}

\draft % marks overfull lines with a black rule on the right

\begin{document}

%\preprint{}

\title{Discrete Electron Emission} %Title of paper



\author{A. Jónsson}
\author{K. Torfason}
\author{A. Manolescu}
\author{Á. Valfells}
 \email[]{av@ru.is}
%\thanks{}
\affiliation{Department of Engineering \\
Reykjavík University \\
Menntavegi 1, 102 Reykjavík, Iceland}


%\date{\today}

\begin{abstract}
Analysis of space-charge effects on electron emission typically makes some assumption of continuity and smoothness, whether this is continuity of charge as in the classical derivation of the Child-Langmuir current, or the mean-field approximation used in particle-in-cell simulations. However, when studying the physics of electron emission and propagation at the mesoscale it becomes necessary to consider the discrete nature of electronic charge to account for the space-charge effect of each individual point charge. In this paper we give an extensive analysis of some previous work on the distribution of electrons under space-charge limited conditions\cite{Gunnarsson_2021, Jonsson_2015}. We examine the spacing of electrons as they are emitted from a planar surface, We present simplified models for analysis of such conditions to derive scaling laws for emission and compare them to computer simulations.    
\end{abstract}

\pacs{}% insert suggested PACS numbers in braces on next line

\maketitle %\maketitle must follow title, authors, abstract and \pacs

\section{Introduction} \label{Introduction}
The effects of space-charge on electron emission have been the subject of intense study for over a century~\cite{Child_1911, Langmuir_1913, Hundred_years_paper, Zhang_2021}. From considerations of one dimensional space-charge limited flow in a planar diode, the theoretical framework has been extended to include various important features, e.g. different geometries~\cite{Luginsland_1996a,Lau_2001,Zhu_2013,Harsha_2022}, finite emission energy~\cite{Jaffe_1944}, relativistic diodes~\cite{Jory_1969}, quantum effects~\cite{Lau_91, Ang_2003}, inhomogeneous cathodes~\cite{Zubair_2016, Chernin_2020, Sitek_2021a}, collisional diodes\cite{Benilov_2009,Darr_2019}, and space-charge influenced field emission~\cite{Barbour_1953,Lau_1994,Feng_2006, Torfason_2015,Jensen_2015}. The bulk of the analysis has been carried out under the assumption that the space-charge forms a continuum. Simulations have primarily been carried out using particle-in-cell (PIC) codes which are based on using aggregate macro-particles and solution of the Poisson equation with a mean-field approximation \cite{Verboncoeur_1995, Verboncoeur_2005}. For quantum effects, solution of the Schrödinger equation for the appropriate system has been used %\cite{Zhang_2016,Zhang_2021}.

This work has proven to be quite useful and accurately describes space-charge effects over a wide parameter range. However, there is a mesoscopic regime where one must consider the discrete nature of electrons, effectively looking at them as interacting point particles. Considering that electron density is generally highest, and their kinetic energy the lowest, near the point of emission, one may anticipate that discrete particle effects such as scattering will be important in that region. This suggests that for a length-scale between the thermal de Broglie wavelength and the electron Debye length in the immediate vicinity of the cathode, it will matter to look at electrons as individual point particles. If one also considers that a single emitted electron will block emission of another electron in some immediate region on the cathode, then it is clear that there must be some minimum spacing between adjacently emitted electrons, or a sort of Coulomb hole surrounding each emitted electron. Thus one must consider discrete particle effects in emitters that have a characteristic dimension comparable to this minimal spacing. In fact, the smallest emitting structures can act as point emitters~
\cite{Gunnarsson_2021}- a regime that has intriguing possible applications for single-electron photoemission that can be used for advanced electron microscopy~\cite{Gevorkyan_2025, Gordon_2025}. Finally, it has been shown that for cathodes with a characteristic feature length, such as the pitch in a field-emitter array, discrete particle effects will be important within a distance from the cathode equivalent to this feature length~\cite{Jensen_2015}.

In this paper we will present and analyze some simple models for space-charge effects from discrete electrons to predict spatial distribution of electrons and derive scaling laws for space-charge limited emission in different regimes. We also present results of simulations for comparison, and compare our results to previous work.

%\label{}
%\subsection{}
%\subsubsection{}

\section{Discrete Space-Charge Models} \label{Discrete Space-Charge Models}

\subsection{General Considerations} \label{General Considerations}

Let us consider a system, consisting of an infinite conductor in the region, $z \leq 0$, subject to a uniform electric field, $-E_0 \hat{\textbf{z}}$, above it and a single electron located at $(r,\varphi,z) = (0,0,\xi)$. The electric field at the surface of the conductor is then found to be $-E_s \hat{\textbf{z}}$ where
\begin{equation}
%\left\{
 E_s=E_0 - \frac{q}{2\pi\varepsilon_0} \frac{\xi}{(r^2+\xi^2)^\frac{3}{2}} \ .
%\right\}.
\label{eq:SurfaceField}
\end{equation}
It is then readily apparent that the electric field at the surface point located directly under the electron, i.~e. at $r=0$ and $z=0$, reverses sign when the electron is at an elevation $\xi_*$ which is given by
\begin{equation}
%\left\{
 \xi_* = \sqrt{\frac{q}{2\pi\varepsilon_0E_0}} \ .
%\right\}.
\label{eq:CharacterLength}
\end{equation}
We will refer to $\xi_*$ as the \textit{critical length} for this system. It will be used substantively as a normalization parameter in this paper. Furthermore, from the same Eq.\ (\ref{eq:SurfaceField}), one can also find that, if $\xi<\xi_*$, the orientation of the total electric field on the conductor surface reverses inside a finite circular area, which has a maximum radius $r_d = (\frac{4}{27})^\frac{1}{4}\xi_* \approx 0.62\xi_*$, corresponding to $\xi = (\frac{1}{3})^\frac{3}{4}\xi_* \approx 0.44\xi_*$. Inside this disk the total field is dominated by the field of the electron, which is positive (considering $q=-e$) and outside this disk by the negative uniform field. This field configuration indicates that it is very unlikely for two electrons to be emitted nearly simultaneously within a distance of $r_d$ from each other.

Another way of looking at this is to bear in mind the relation between the surface charge density and the applied field, which in the absence of space-charge is given by $\sigma=\varepsilon_0E_0$. Thus if one wanted to form a single electron, of total charge $q$, from the surface charge, an area of radius, $R_*=\sqrt{\frac{q}{\pi\varepsilon_0E_0}}=\sqrt{2}\xi_*$, is needed. Clearly, $r_d \neq R_*$ but they are of the same order of magnitude. Therefore we expect that the limiting spacing between adjacent electrons will be comparable to these values as well.  In light of this, the following numerical relation proves useful:
\begin{equation}
%\left\{
 \xi_* = \frac{54}{\sqrt{E_0}} \ ,
%\right\}.
\label{eq:CharacterLengthNum}
\end{equation}
with $\xi_*$ measured in nm and $E_0$ in MV/m.

In the analysis of the cases to follow, we will be using equations for electric fields, potential, etc. that can be quite cumbersome. For clarity of presentation we introduce the following normalized parameters: $\overline{\xi}=\frac{\xi}{\xi_*}$ for distance, $\overline{t}=t\sqrt{\frac{qE_0}{m\xi_*}}$ for time, $\overline{E}=\frac{E}{E_0}$ for electric field, $\overline{\phi}=\frac{\phi}{E_0\xi_*}$ for electric potential, and $\dot{\overline{\xi}}=\sqrt{\frac{m}{qE_0\xi_*}}\dot{\xi}$ for velocity.

\subsection{Point emitter in a uniform field} \label{Point emitter in a uniform field}

Motivated by previous work on electron emission from a point emitter~\cite{Gunnarsson_2021}, we will now proceed to derive an equation for the maximum current that may be drawn from a point emitter. We do this by looking at the shortest possible interval for sequential emission of electrons from a fixed point on a planar conductor such as that described in the preceding discussion. We assume that electrons are launched at an elevation above this point which corresponds to the maximum of the surface barrier, with a fixed velocity normal to the surface. The first electron is launched at time $\overline{t} = 0$. We denote its position, velocity, electric field, and electric potential at its location, by $\overline{\xi}_1$, $\dot{\overline{\xi}}_1$, $\overline{E}_1$, and $\overline{\phi}_1$, respectively. Corresponding values for the $n$-th electron launched would make use of the subscript $n$. For a single emitted electron we have the equation of motion
\begin{equation}
%\left\{
 \ddot{\overline{\xi}}_1 = 1-\frac{1}{8\overline{\xi}_1^2} \ ,
%\right\}.
\label{eq:SingleElectronMotion}
\end{equation}
with the initial conditions $\overline{\xi}_1(0) = \frac{1}{\sqrt{8}}$ and $\dot{\overline{\xi}}_1(0) = \overline{v}_0$. 

The next step is to include the coupling between the first and second electron. To do this we begin by calculating the acceleration of the second emitted electron. This yields
\begin{equation}
%\left\{
 \ddot{\overline{{\xi}}}_2 = 1-\frac{1}{2}\left(\frac{1}{\left(2\overline{\xi}_2\right)^2}+\frac{1}{\left(\overline{\xi}_1-\overline{\xi}_2\right)^2}+\frac{1}{\left(\overline{\xi}_1+\overline{\xi}_2\right)^2}\right) \ .
%\right\}.
\label{eq:SecondElectronMotion}
\end{equation}
The first term within the parentheses is due to the image charge of the second electron emitted, the second term is due to interaction between the two electrons, and the third term is due to the image charge of the first electron emitted. We next assume as the condition for launch of the second electron that there exists a pair of real values for $\xi_1$ and $\xi_2$ such that $0<\overline{\xi}_2<\overline{\xi}_1$ and $\ddot{\overline{\xi}}_2=0$. In other words that the electric potential $\phi_2$ has a maximum. Setting the acceleration to zero, Eq.\ (\ref{eq:SecondElectronMotion}) may be recast as 
\begin{equation}
%\left\{
 8X_2^3-\left(16\overline{\xi}_1^2+9\right)X_2^2 +\left(8\overline{\xi}_1^4-6\overline{\xi}_1^2\right)X_2-\overline{\xi}_1^4 \ ,
%\right\}.
\label{eq:SecondElectronRecast}
\end{equation}
with $X=\overline{\xi}_2^2$. It can be shown that Eq.\ (\ref{eq:SecondElectronRecast}) has three distinct real roots in $X$, but to find them as a function of $\overline{\xi}_1$ is laborious and not illustrative. Thus, we solve it numerically for the smallest value of $\overline{\xi}_1$ such that $0<\overline{\xi}_2<\overline{\xi}_1$. This solution corresponds to a time $\overline{t}_e\approx 1.580$ and positions $\overline{\xi}_1\left(\overline{t}_e\right)\approx1.557$ and $\overline{\xi}_2\left(\overline{t}_e\right)\approx0.612$.

From this one can make a first approximation of the maximum current, namely $I_{max}\approx \frac{q}{t_e}$. However, we realize that with a larger number of electrons emitted the actual interval between emission events, $\tau$, must be greater than $t_e$ due to the increased space-charge effects. Nonetheless, from these considerations it is apparent that the space-charge limiting current from a point emitter will take the form
\begin{equation}
%\left\{
 I_{max} = \frac{q}{\tau}=\frac{q}{\sqrt{\frac{m\xi_*}{qE_0}}\overline{\tau}}=\frac{1}{\overline{\tau}}\left(\frac{2\pi\varepsilon_0q^5}{m^2}\right)^\frac{1}{4}E_0^\frac{3}{4} \ .
%\right\}.
\label{eq:PointEmitter}
\end{equation}
The value of $\overline{\tau}$ depends on the number of electrons in the train of charge and the emission velocity but, as will be shown in the simulation results, for a wide range of parameters $1 \lessapprox \overline{\tau} \lessapprox2$. This scaling with the electric field is completely different from that which is commonly observed for space-charge limited current from a finite emitting area. This new scaling has been observed before for point emission~\cite{Gunnarsson_2021,Gevorkyan_2025}. The transition from the conventional Child-Langmuir scaling of $I \propto E_0^{3/2}$ to the point emitter scaling of $I \propto E_0^{3/4}$ has previously been shown to be smooth with diminishing emitter area~\cite{Gunnarsson_2021}.

\subsection{Discrete Sheet of Charge} \label{Discrete Sheet of Charge}
Let us now examine the following configuration in a Cartesian system: As before we assume that a perfect conductor fills the space, $z\leq0$ and there is an electric field $-E_0 \hat{\textbf{z}}$ above the conductor. Next, an infinite number of electrons is placed on a rectangular grid in the plane $z=\xi > 0$. The distance between adjacent electrons is $\Delta x, \Delta y=l$. We refer to this distance as the \textit{pitch} of the array (the lattice constant). Each rectangular area bounded by four adjacent electrons is referred to as the (unit) \textit{cell}.

\begin{figure}[t]
%    \centering
    \includegraphics[width=\linewidth]{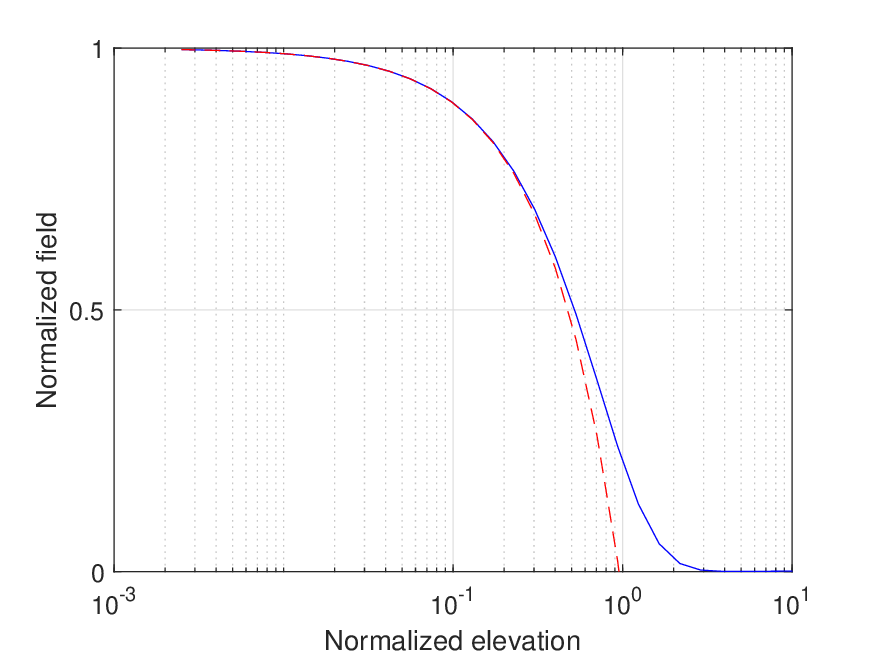}
    \caption{Surface field, $\overline{E}_{2mD}$, from Eq. (~\ref{eq:SheetEmitterNorm}) as a function of elevation for a discrete sheet of particles. $\overline{l}=\sqrt{2\pi}$. $\overline{E}_{m2D}\approx1-\overline{\xi}\frac{16.5172}{\overline{l}^3}$ shown as dashed line.}
    \label{fig:Discrete_sheet}
    % \vspace{-0.5cm}
\end{figure}

We are interested in examining the electric field on the conductor surface directly underneath the center of a cell. This is because it is the location least influenced by the space-charge field from the sheet, and thus most likely to be a candidate site for emission of an electron. The normalized electric field at such a location is $-E_{m2D} \hat{\textbf{z}}$, where
\begin{equation}
%\left\{
\begin{split}
& E_{m2D}  = E_0 - \frac{q}{2\pi\varepsilon_0} \\\\ 
& \times 4\sum_{k=0}^{\infty}\sum_{h=0}^{\infty}\frac{\xi}{\left(\xi^2+\left(h+\frac{1}{2}\right)^2l^2+\left(k+\frac{1}{2}\right)^2l^2\right)^{3/2}} \ .
\end{split}
%\right\}.
\label{eq:SheetEmitter}
\end{equation}
In normalized variables this becomes
\begin{equation}
%\left\{
\begin{split}
& \overline{E}_{m2D}  = 1 - \frac{4}{\overline{\xi}^{2}}\\\\ 
& \times \sum_{k=0}^{\infty}\sum_{h=0}^{\infty}\frac{1}{\left(1+\left(h+\frac{1}{2}\right)^2u^2+\left(k+\frac{1}{2}\right)^2u^2\right)^{3/2}} \ ,
\end{split}
%\right\}.
\label{eq:SheetEmitterNorm}
\end{equation}
with $u=\overline{l}/\overline{\xi}$. The double sum in the preceding equation has two asymptotes, namely $\frac{\pi}{2}u^{-2}$ for $u<<1$ and $4.1293u^{-3}$ for $u>>1$. Thus
\begin{equation}
%\left\{
\overline{E}_{m2D} \approx 1 - \frac{4}{\overline{\xi}^{2}}\frac{\pi}{2u^2}=1-\frac{2\pi}{\overline{l}^2} \ ,
%\right\}.
\label{eq:Small_l_asymptote_p}
\end{equation}
for $\overline{l}<<\overline{\xi}$, and 
\begin{equation}
%\left\{
\overline{E}_{m2D} \approx 1 - \frac{4}{\overline{\xi}^{2}}\frac{4.1293}{u^3}=1-\overline{\xi}\frac{16.5172}{\overline{l}^3} \ ,
%\right\}.
\label{eq:Large_l_asymptote_p}
\end{equation}
for $\overline{l}>>\overline{\xi}$. Figure~\ref{fig:Discrete_sheet} shows $E_{m2D}$ as a function of elevation for $\overline{l}=\sqrt{2\pi}$ as well as the asymptote given in  Eq.\ (\ref{eq:Large_l_asymptote_p}).

From Eq.\ (\ref{eq:Small_l_asymptote_p}) we see that when the discrete sheet is elevated high above the conductor, the surface electric field becomes essentially the same as would be obtained using a continuous charge model for the sheet. It also shows that the smallest allowable pitch such that the surface electric field at the point of interest is never oriented so as to oppose admission is $\overline{l}=\sqrt{2\pi}$ which corresponds to an average charge density of the sheet equal to $\sigma=q/l^2=q/\left(2\pi\xi_*^2\right)=\varepsilon_0E_0$ which is simply the surface charge density of the conductor exposed to the applied field $E_0$. This also means that the minimum value of the pitch is $l = \sqrt{\pi}R_*$ and thus conforms quite well to the smallest spacing between emitted electrons obtained from the earlier estimate using $\pi R_*^2=q$. That the discrete sheet model recovers the continuous sheet results in such a manner also means that the space charge limited current will scale as $E_0^{3/2}$ as predicted by the capacitive derivation of the Child-Langmuir law~\cite{Umstattd_2005}. 
\subsection{Discrete String of Charge} \label{Discrete String of Charge}
\begin{figure}[tb]
% \centering
 \begin{subfigure}[t]{0.8\linewidth}
   \includegraphics[width=\linewidth,trim={0.2cm 0cm 1.5cm 1cm},clip]{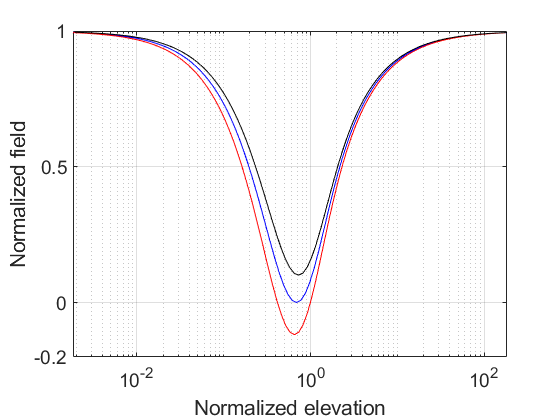}
\caption{String of electrons with spacing of $\overline{l}=1.734$ (red), $\overline{l}=1.834$ (blue), and $\overline{l}=1.934$ (black).}
   \label{fig:Line_emitter_field_a}
 \end{subfigure}\hfill
 \begin{subfigure}[t]{0.8\linewidth}
   \includegraphics[width=\linewidth,trim={0.2cm 0cm 1.5cm 1cm},clip]{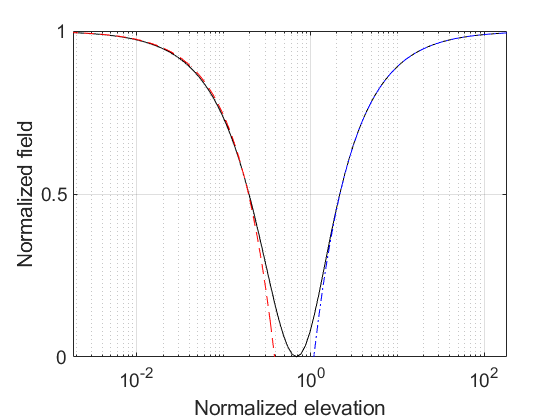}
\caption{$\overline{l}=1.834$ (black), asymptotes $\overline{E} \approx 1-\frac{2}{\overline{\xi}\overline{l}}$ (blue dash-dot) and $\overline{E} \approx 1-\frac{16\overline{\xi}}{\overline{l}^3}$ (red dash).}
   \label{fig:Line_emitter_field_b}
 \end{subfigure}\hfill
\caption{The surface field, $\overline{E}_{m1D}$, from Eq. (~\ref{eq:StringEmitterNorm}) as a function of $\overline{\xi}$, for discrete string of particles}.
 \label{fig:Line_emitter_field}
  \vspace{-0.75cm}
\end{figure}

We now turn our attention to a model that is similar to the one in the preceding section, except that instead of a discrete sheet of electrons, we restrict all of the electrons to be on the line described by $y=0$ and $z=\xi$. The number of electrons is infinite and they are spaced with a uniform pitch $l$. %{Physically, this could be representative of emission from a sharp ridge, grain boundary or even the rim of a straight carbon-nanotube fiber.}
We wish to find the electric field at the conductor surface directly below the mid-point between two adjacent electrons. Again, this is because it corresponds to the weakest space-charge effect on the part of the surface that lies directly under the string of electrons. At these points the electric field on the conductor surface is $-E_{m1D} \hat{\textbf{z}}$. In normalized terms we have
\begin{equation}
%\left\{
\overline{E}_{m1D} = 1 - \frac{2}{\overline{\xi}^{2}}\sum_{k=0}^{\infty}\frac{1}{\left(1+\left(k+\frac{1}{2}\right)^2u^2\right)^{3/2}} \ ,
%\right\}.
\label{eq:StringEmitterNorm}
\end{equation}
where $u=\overline{l}/\overline{\xi}$ as before. The sum has the asymptote $u^{-1}$ as $u$ tends to zero and the asymptote $8u^{-3}$ as $u$ tends to infinity. From this we obtain
\begin{equation}
%\left\{
\overline{E}_{m1D} \approx 1 - \frac{2}{\overline{\xi}\overline{l}} \ ,
%\right\}.
\label{eq:StringEmitter_small_l}
\end{equation}
for $\overline{l}<<\overline{\xi}$, and
\begin{equation}
%\left\{
\overline{E}_{m1D} \approx 1 - \frac{16\overline{\xi}}{\overline{l}^3} \ ,
%\right\}.
\label{eq:StringEmitter_large_l}
\end{equation}
for $\overline{l}>>\overline{\xi}$. We now look for the pitch that corresponds to space-charge limited emission. In other words, we want to space the electrons as closely as possible so that $E_{m1D}$ in non-negative for all values of $\xi$. This is solved numerically and yields the pitch, $\overline{l}_{m1D} = 1.834$. Figure ~\ref{fig:Line_emitter_field} shows $E_{m1D}$ as a function of elevation for different values of pitch.

\section{Simulations} \label{Simulations}
\subsection{Molecular dynamics simulations} \label{Molecular dynamics simulations}
We have used our molecular dynamics code RUMDEED \cite{Torfason_2015,Torfason_2021,Sitek_2021a,Sitek_2021b} to perform simulations of different scenarios for comparison with the analytic models described in the previous section. A rough overview of the code structure follows:

\begin{enumerate}
  \item Emission for the current time-step is calculated for a specified emission mechanism, taking into account electric field at cathode surface due to external field and space-charge from individual emitted electrons and their image-charge partners. Electrons are placed in system and space-charge updated with every added electron.  
  \item Once all electrons have been emitted, the force on each individual electron is calculated from external field and direct Coulomb interaction with every other free electron and image charge partner in the system. New position and velocity assigned.
  \item Current is calculated and electrons that cross the system boundary are removed.
  \item The time is advanced and step 1 is repeated. 
\end{enumerate}

As the code does not make use of a grid to solve the Poisson equation, there is no mean-field approximation and discrete particle effects are not obscured. However, the temporal evolution is discrete with a fixed time-step. Calculation of the local electric field at the cathode surface takes into account space-charge contribution from individual electrons and their image charge partners, thus also preserving the effects of granularity on emission. There are a number of self-consistent emission modules, but for the following simulations we will use enforced space-charge limited emission.

\subsubsection{Space-charge limited emission from a ring} \label{Space-charge limited emission from an ring}
We begin by simulating space-charge limited emission from a ring embedded in the cathode. Its outer radius is $R = 50$ nm and the gap spacing of the diode is $D=1000$ nm. The gap potential, $V$, is varied. We choose the width of the ring $W = R_*$, which scales as $V^{-1/2}$. By using this value of $W$ and having $R>>R_*$ we are effectively mimicking emission from a string in the sense that the width of the ring cannot accommodate two emission sites within a small interval of the polar angle, $\delta \phi$. Similarly the large radius has the effect that the emission site at polar angle $\phi_1$ is not influenced by electrons emitted at a site at $\phi_1 + \pi$. The time-step is $\Delta t=1.25$ fs.

\begin{figure}[t]
 \centering
 \begin{subfigure}[t]{0.75\linewidth}
   \includegraphics[width=\linewidth,clip]{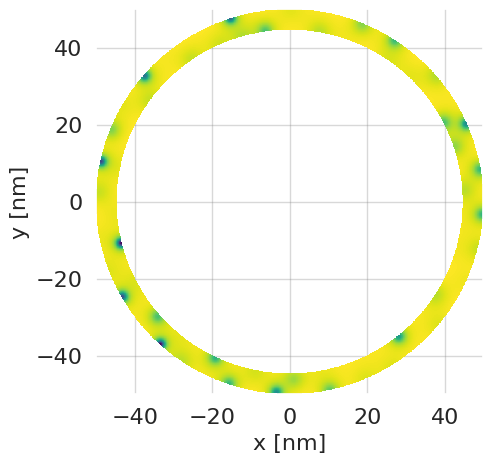}
\caption{Electric field on ring surface at arbitrary time $t_a$}
   \label{fig:Field_plot_1}
 \end{subfigure}\hfill
 \begin{subfigure}[t]{0.75\linewidth}
   \includegraphics[width=\linewidth,clip]{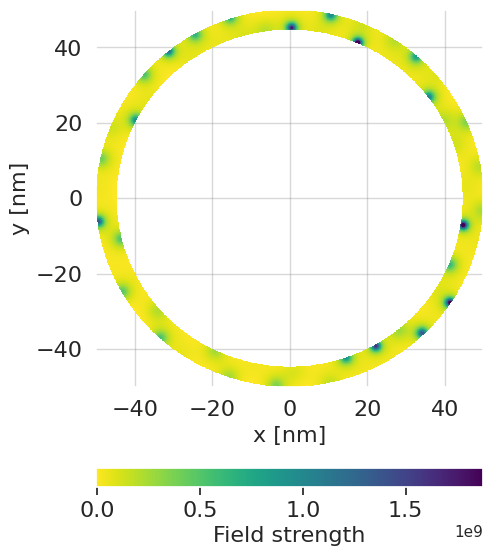}
\caption{Electric field on ring surface at time $t_a+62.5$ fs.}
   \label{fig:Field_plot_2}
 \end{subfigure}\hfill
\caption{The magnitude of the surface electric field on the emitting ring at two instances in time. $D = 1000$ nm, $V = 200$ V.
    The bottom picture shows the situation after 62.5 fs have elapsed from the top picture is taken. Note the signature effects of newly emitted electrons.}.
 \label{fig:Field_plot}
  \vspace{-0.75cm}
\end{figure}

Figure~\ref{fig:Field_plot} shows the magnitude of the electric field on the ring surface for two instances in time. We see that the electric field is strongest (and unfavorably oriented for further emission) at locations under newly emitted electrons. We also see that after time interval of $\Delta t = 62.5$ fs, the strength of the electric field at the previous emission sites has diminished greatly as those electrons propagate away from the cathode, and newly emitted electrons appear at locations removed from the prior ones. 

% \begin{figure}[h]
%     \centering
%     \includegraphics[width=\linewidth]{Figures/Field_plot.png}
%     \caption{The magnitude of the surface electric field on the emitting ring at two instances in time. $D = 1000$ nm, $V = 200$ V.
%     The bottom picture shows the situation 6.25 fs the top picture. Note the signature effects of newly emitted electrons.}
%     \label{fig:Field_plot}
% \end{figure}

\begin{figure}[t]
%    \centering
    \includegraphics[width=\linewidth]{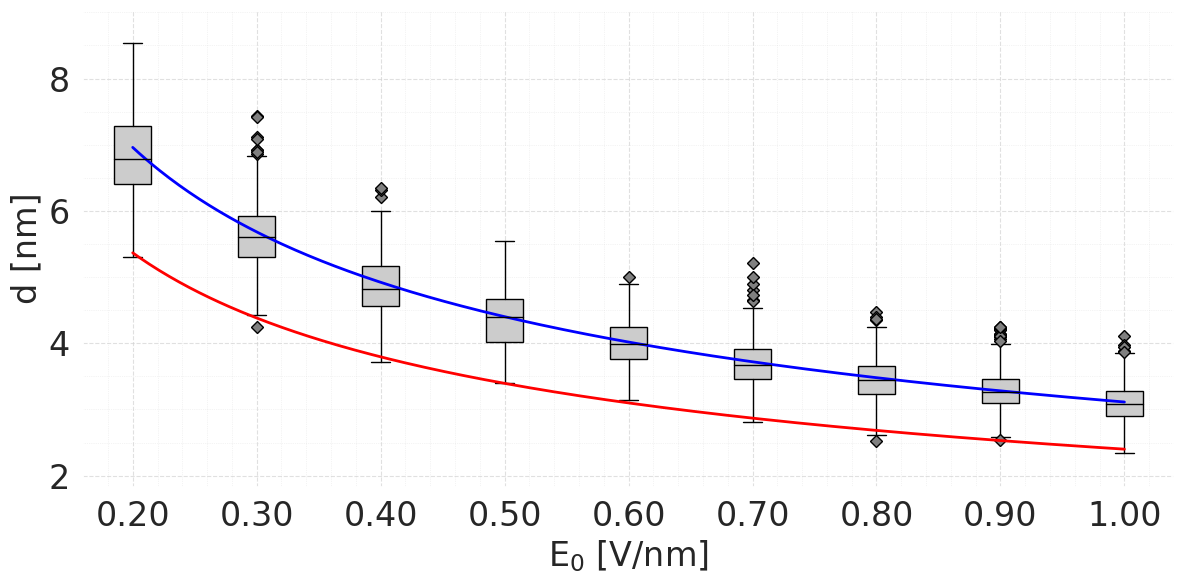}
    \caption{Distribution of separation between adjacent electrons emitted from a ring. Boxplot shows first to third quartile with line at the median. Whiskers extend to 1.5 times the inter-quartile range. Diamonds show statistical outliers. Blue line shows pitch spacing $=1.834\xi_*$ and red line shows $R_*=\sqrt{2}\xi_*$ .}
    \label{fig:Ring_distribution_1}
\end{figure}

\begin{figure}[t]
    \centering
    \includegraphics[width=\linewidth]{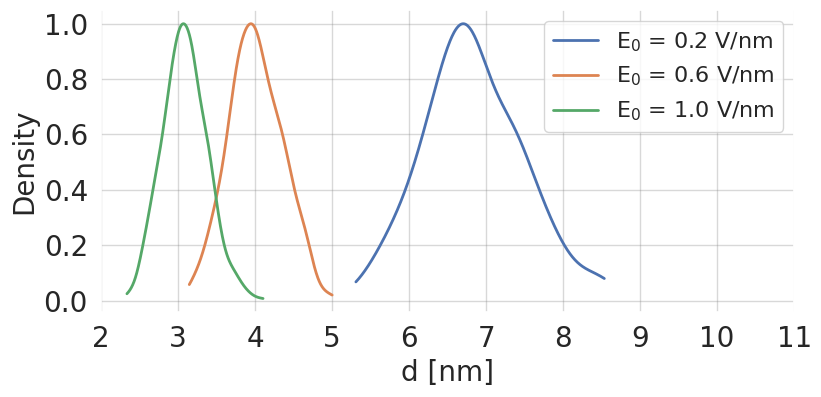}
    \caption{Distribution of distance between adjacent electrons, emitted from the ring, for three different values of applied field.}
    \label{fig:Ring_distribution_2}
\end{figure}

Figure~\ref{fig:Ring_distribution_1} shows how the separation between adjacent electrons in the immediate vicinity of the cathode varies with the strength of the applied field. It is interesting to note that the separation scales as $E^{-1/2}$, that the lower bound of the spacing is set by $R_*$, and that the average spacing between adjacent electrons is described by the normalized parameter $\overline{l}=1.834$ as predicted by the discrete string model described in Section~\ref{Discrete String of Charge}. Figure~\ref{fig:Ring_distribution_2} shows another view of the distribution of electron separation for three different values of the applied field.

Having established that the spacing between electrons emitted from the ring under space-charge limited conditions is on average $l=1.834\xi_*$, it is easy to apply the same line of reasoning that was used to determine the space-charge limited current from a point, to the ring. If we denote the space-charge limited current from the ring as $I_r$, then we can use the estimate $I_r=\frac{qN_e}{\tau}$, where $\tau$ is the interval between successive emissions from a point, $N_e=\frac{2\pi R}{1.834\xi_*}$ is the number of emission sites on the ring at a given time, and $R$ is the radius of the ring. Hence,

\begin{equation}
%\left\{
I_r = \frac{2\pi Rq}{\overline{\tau}1.834\xi_*\sqrt{\frac{m\xi_*}{qE_0}}}=\frac{2\pi R}{1.834\overline\tau}\frac{\left(2\pi q\varepsilon_0\right)^{3/4}}{\sqrt{m}}E_0^{5/4} \,.
%\right\}.
\label{eq:RingCurrent}
\end{equation}

\begin{figure}[t]
    \centering
    \includegraphics[width=\linewidth]{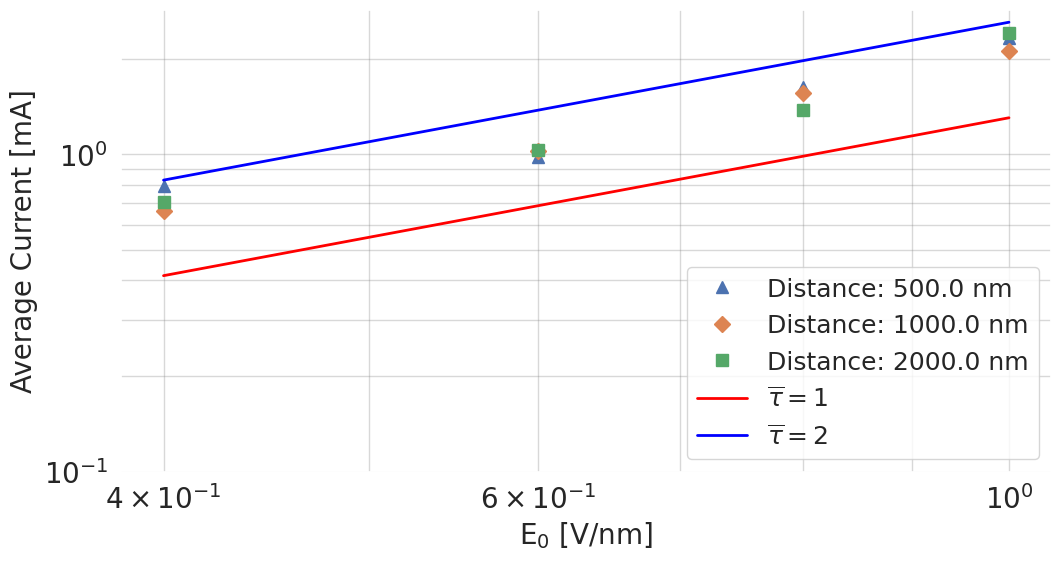}
    \caption{Space charge limited emission from the ring as a function of electrostatic field strength for three different gap spacings: $D = 500$ nm, $D = 1000$ nm, and $D = 2000$ nm. The blue line shows the space-charge limited current calculated from Eq.~(\ref{eq:RingCurrent}) using $\overline{\tau}=1$ and the red line for $\overline{\tau}=2$.}
    \label{fig:Ring_current_plot}
\end{figure}

We will later see that $1 \lessapprox \overline{\tau} \lessapprox2$ in most instances for the string emitter, depending on emission velocity. Figure \ref{fig:Ring_current_plot} shows how the space-charge limited current from the ring scales with the external electric field for three different values of the gap spacing and compared to Eq.~(\ref{eq:RingCurrent}) using $\overline{\tau}=1$ and $\overline{\tau}=2$.

\subsubsection{Space-charge limited emission from a circular patch} \label{Space-charge limited emission from a circular patch}
\begin{figure}[t]
    \centering
    \includegraphics[width=\linewidth]{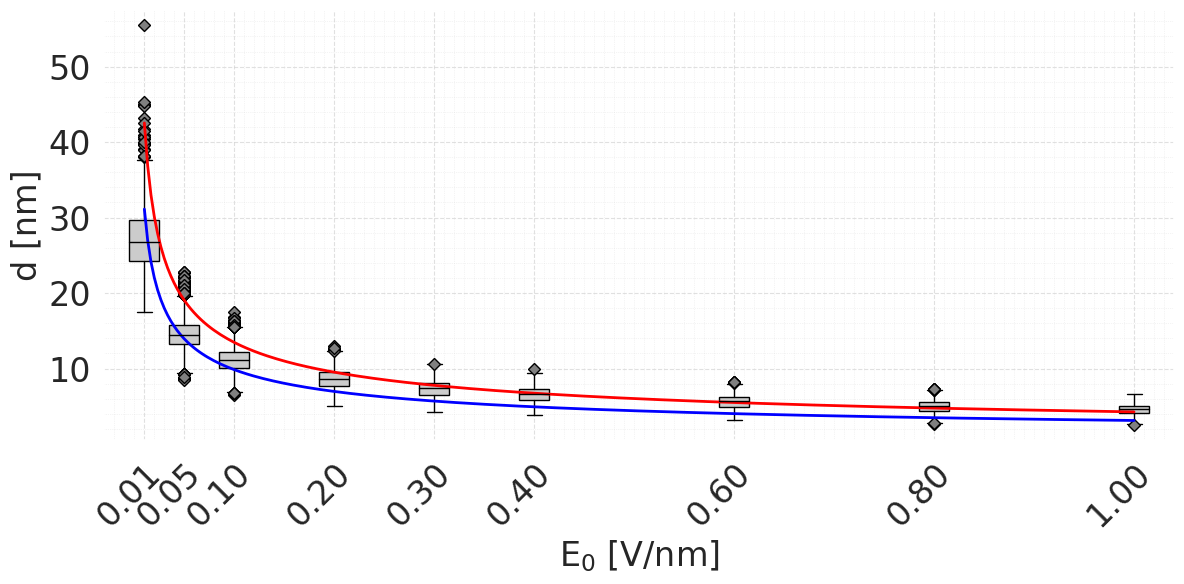}
    \caption{Distribution of spacing between adjacent electrons emitted from a circular patch. The boxplot shows first to third quartile with line at the median. Whiskers extend to 1.5 times the inter-quartile range. Diamonds show statistical outliers. The blue line shows spacing $=1.834\xi_*$ and red line shows $l=\sqrt{2\pi}\xi_*$ .}
    \label{fig:Circle_plot_1}
\end{figure}

\begin{figure}
    \centering
    \includegraphics[width=\linewidth]{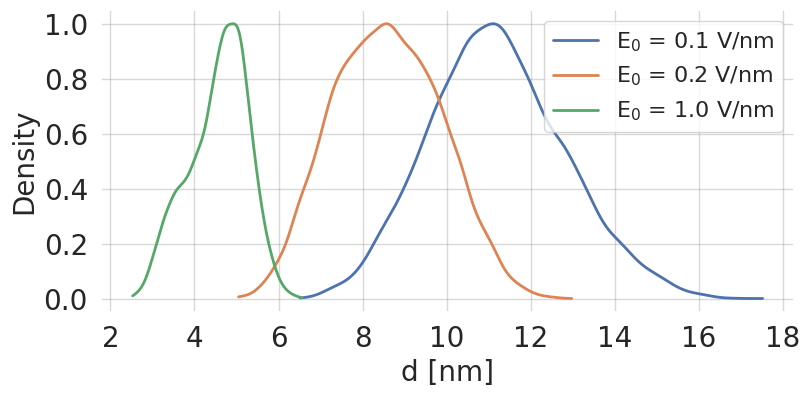}
    \caption{Distribution of distance between adjacent electrons, emitted from the circular patch, for three different values of applied field. Note how the "shoulder" of the distribution varies with the applied field.}
    \label{fig:Circle_plot_2}
\end{figure}

We next simulate space-charge limited emission from a circular patch embedded in the cathode of a planar diode. The radius of the patch is $R=50$ nm and the gap spacing of the diode is $D=1000$ nm, the potential, $V$ is varied. The time-step is $\Delta t = 1.25$ fs. Figure~\ref{fig:Circle_plot_1} shows the distribution of spacing between nearest neighbors. Note the shift in average spacing as the applied field grows. For low values of the applied field the average spacing tends to that of $\overline{l}=1.834$ as for a ring emitter, whereas for higher values of the applied field it tends towards $\overline{l}=\sqrt{2\pi}$ as for the sheet. This is a manifestation of a previously observed effect that space-charge limited current density is greater at the edge of a finite emitter~\cite{Luginsland_2002} and that the proportion of the total charge coming from edge increases with decreasing field strength to a degree that up to 90 percent of the total current may come from the edge~\cite{Gunnarsson_2021}. Figure~\ref{fig:Circle_plot_2} shows the distribution of distances between nearest those adjacent electrons for low and high field strengths. The skewing of the distribution due to enhanced contribution from the emitter rim is readily apparent.

%\subsubsection{Field emission from an ring}
%When considering field emission one should bear in mind that the space-charge limit will not be reached since a finite electric field is necessary for emission. Hence we talk about space-charge influenced field emission, an equilibrium condition determined by space-charge suppression of the surface electric field. We anticipate that the electron spacing will be greater than under space-charge limited conditions, but are primarily interested in seeing how the spacing is affected by the applied electric field. A thorough analysis is beyond the scope of this paper, but we present some indicative results for space-charge influenced field emission.
\subsection{Simulations of vanishing emitter area} \label{Simulations of vanishing emitter area}
To get a better understanding of point emission and emission from an infinitely thin line, we conducted a number of numerical experiments. From these we could determine the value of the interval, $\overline{\tau}$, between sequential electron emissions from a point or line for different values of emission velocity.

\begin{figure}[t]
 \centering
 \begin{subfigure}[t]{\linewidth}
   \includegraphics[width=\linewidth,clip]{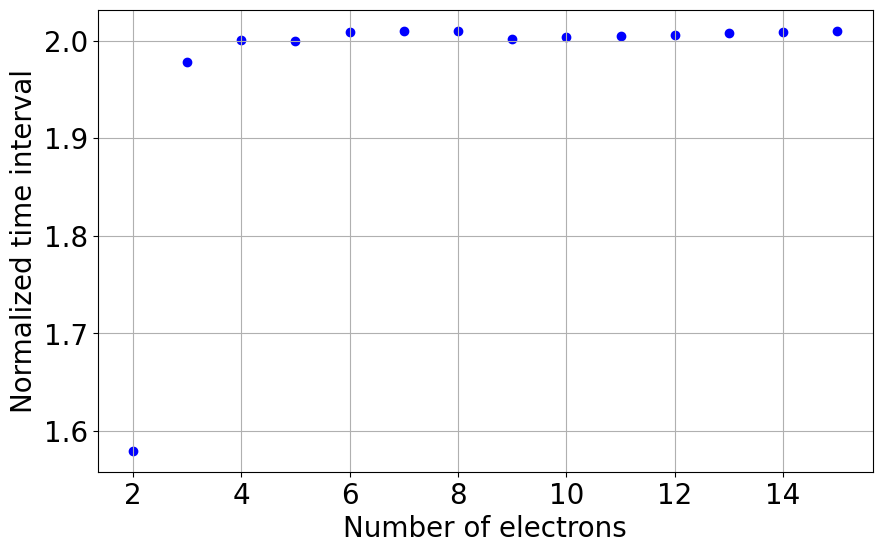}
\caption{Variation of $\overline{\tau}$ with the number of electrons emitted. Initial velocity is $\overline{v}_0=10^{-3}$.}
   \label{fig:Point_plot1}
 \end{subfigure}\hfill
 \begin{subfigure}[t]{\linewidth}
   \includegraphics[width=\linewidth,clip]{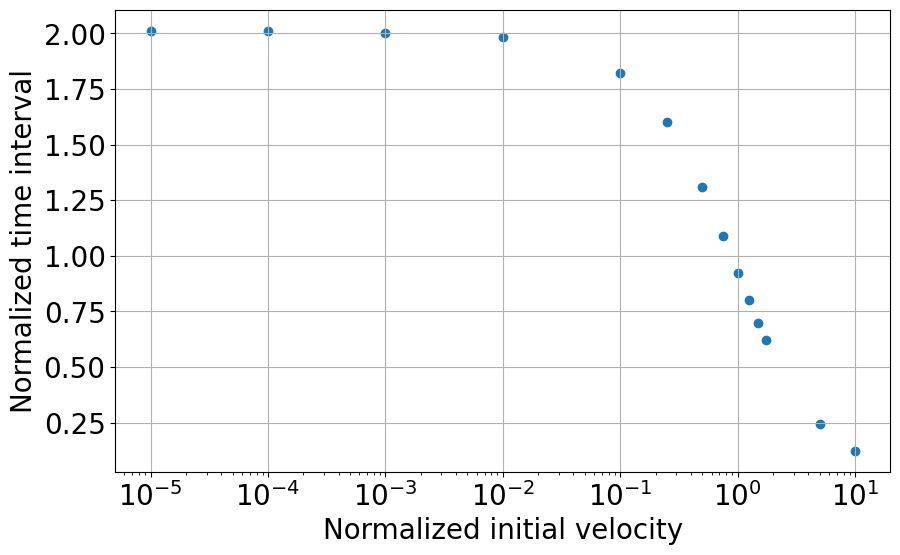}
\caption{Variation of $\overline{\tau}$ with $\overline{v}_0$ for 10 emitted electrons.}
   \label{fig:Point_plot2}
 \end{subfigure}\hfill
\caption{Normalized interval, $\overline{\tau}$, between successive emissions from a point.}
 \label{fig:Point_plot}
%\vspace{-0.75cm}
\end{figure}

We consider sequential emission and propagation of electrons from a single point. Let $\overline{\xi}_n$ denote the normalized elevation of the $n$-th emitted electron above the cathode which is in the $\overline{\xi}=0$ plane. Upon emission, an electron is placed at an elevation where it may first experience an accelerating field, and given a normalized velocity of $\overline{v}_0$ perpendicular to the cathode. We take the image charge of each electron into account. Thus, the first electron will be placed such that at time $\overline{t}_1=0$ it is at $\overline{\xi}\left(\overline{t}_1\right)=1/\sqrt{8}$ as per equation Eq.\ (\ref{eq:SingleElectronMotion}). This electron propagates under the influence of the applied field and that of its image charge. We then advance time and the location of the first electron until Eq.\ (\ref{eq:SecondElectronMotion}) has a solution with $\ddot{\overline{{\xi}}}_2=0$ and $0<\overline{{\xi}}_2\left(\overline{t}_2\right)<\overline{{\xi}}_1\left(\overline{t}_2\right)$. Thus, at time $\overline{t}_2$ the second electron is placed at $\overline{{\xi}}_2\left(\overline{t}_2\right)$ with an initial velocity $\overline{v}_0$. We record the time interval $\overline{\tau}=\overline{t}_2-\overline{t}_1$ and proceed to advance electrons 1 and 2 under the influence of the applied field, their mutual interaction, and interaction with the 2 image charges. We continue adding electrons in this fashion until we see a stabilization in the interval between emitted electrons. Figure~\ref{fig:Point_plot} shows the evolution of $\overline{\tau}$ as a function of emission velocity and number of electrons emitted. From this we see that for most practical situations, $1 \lessapprox \overline{\tau} \lessapprox2$.

\begin{figure}[t]
    \centering
    \includegraphics[width=\linewidth]{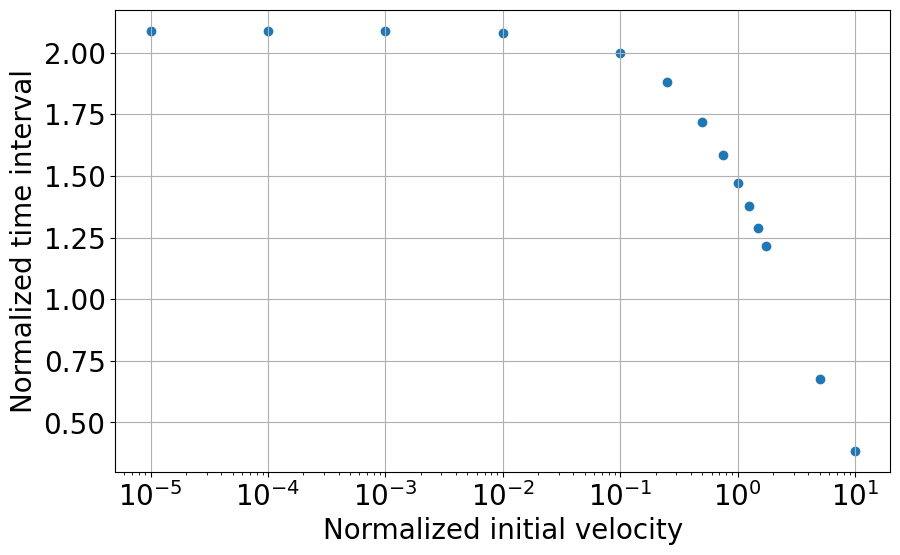}
    \caption{Normalized interval, $\overline{\tau}$, between successive emissions from a line emitter. There are 60 electrons per string emitted, and the number of strings is 8.}
    \label{fig:Line_plot}
\end{figure}

A similar procedure can be carried out for an infinite string of equally spaced electrons being emitted from a line of infinitesimal thickness, essentially the model described in Section \ref{Discrete String of Charge}. The difference is that the emission points for successive generations of electrons are shifted by $\overline{l}/2$ since it is directly below the midpoint of two neighboring electrons that the cathode opens up for emission first. Figure~\ref{fig:Line_plot} shows the evolution of $\overline{\tau}$ for the string of charge. As the number of emitter sites per length of string, $L$, is $N_e=\frac{L}{1.834\xi_*}$ we find that the linear current density is $\frac{qN_e}{\tau}$ which scales as $E_0^{5/4}$.

It is good to bear in mind that the normalized velocity relates to the initial velocity as $\overline{v}_0=\sqrt{\frac{m}{qE_0\xi_*}}v_0$ from which a useful estimate of the normalized emission velocity can be obtained: 
\begin{equation}
%\left\{
\overline{v}_0 \approx 6.1\frac{\mathscr{E}^{1/2}}{E_0^{1/4}} \,  \, .
%\right\}
\label{eq:EmissionVelocity}
\end{equation}
Here $\mathscr{E}$ is the emission energy in eV and the electrostatic field $E_0$ is measured in MV/m.


\section{Conclusions} \label{Conclusions}

We constructed simple models to estimate discrete particle effects for space-charge limited emission for a point emitter, line emitter and plane emitter embedded in a planar cathode subject to an applied electric field of strength $E_0$. From these models we obtain the average distance between electrons that are emitted, under space-charge limited conditions, within a short time period $\Delta t << \tau$, where $\tau \approx \left(\frac{m^2}{\pi \varepsilon_0 q E_0^3}\right)^{3/4}$ is the characteristic time between successive emission from a given location. Simulations confirm the validity of the average spacing obtained from the analytic models. Furthermore, we have identified a characteristic scale, the \textit{critical length} $\xi_* = \sqrt{\frac{q}{2\pi\varepsilon_0E_0}}$ that is present in  different scalings of the space-charge limited current. This length represents the minimum elevation of a just-emitted electron which yields a zero total electric field at the cathode surface. For an emitting patch of finite area, where the largest characteristic dimension is smaller than the critical length, we may treat it as a point emitter and show that the space-charge limited current scales as $E_0^{3/4}$. For a finite emitting surface, with one characteristic dimension greater than the critical length and the other smaller than it, we may treat it as a line (or string) emitter, and show that the space-charge limited current scales as $E_0^{5/4}$. If both dimensions are much greater than the critical length we recover a scaling of the space-charge limited current that corresponds to the classic Child-Langmuir law, namely $E_0^{3/2}$.

This work pertains to a planar configuration and uniform field. It is of interest to look at different geometries, such as a sharp tip or sphere where the field becomes nonuniform and deviations from the presented scaling laws are expected.

\section{Acknowledgements}
This work was supported by the Air Force Office of Scientific Research under Award no. FA8655-23-7003.

\section{Data Availability}
The data that supports this study are available from the corresponding author upon reasonable request.

\section{Author Declarations}

\subsection{Conflicts of interest}
The authors have no conflicts to disclose.

% If in two-column mode, this environment will change to single-column format so that long equations can be displayed. 
%\begin{widetext}
%$$\mbox{put long equation here}$$
%\end{widetext}

% Figures should be put into the text as floats. 
% Use the graphics or graphicx packages (distributed with LaTeX2e).
% Put the label that you will use with \ref{} command in the braces of the \label{} command.
%
% \begin{figure}
% \end{figure}

% \begin{table}
% \end{tabular}
% \end{table}

%\begin{acknowledgments}
% Put your acknowledgments here.
%\end{acknowledgments}

\bibliographystyle{apsrev4-2}
\bibliography{references}

\end{document}